\documentclass[twocolumn]{aastex63} 
\usepackage{csvsimple}
\usepackage{multirow,bigdelim}

\usepackage{amsmath}

\newcommand{\msun}{\mbox{$M_\odot$}}

\newcommand{\mearth}{M$_\oplus$}

\newcommand{\kms}{\mbox{km s$^{-1}$}}

\newcommand{\om}{`Oumuamua}
\newcommand{\htwo}{H$_2$}
\newcommand{\ntwo}{N$_2$}
\usepackage{colortbl}
\usepackage{lineno}
\definecolor{tablegray}{rgb}{0.89, 0.89, 0.89}

\begin{document}

\title{Constraints on the Occurrence of `Oumuamua-Like Objects}
\correspondingauthor{W. Garrett Levine}

\email{garrett.levine@yale.edu}

\author[0000-0002-1422-4430]{W. Garrett Levine}
\affil{Dept. of Astronomy, Yale University, 52 Hillhouse, New Haven, CT 06511, USA}

\author[0000-0001-9749-6150]{Samuel H. C. Cabot}
\affil{Dept. of Astronomy, Yale University, 52 Hillhouse, New Haven, CT 06511, USA}

\author[0000-0002-0726-6480]{Darryl Seligman}
\affil{Dept. of the Geophysical Sciences, University of Chicago, Chicago, IL 60637}

\author[0000-0002-3253-2621]{Gregory Laughlin}
\affil{Dept. of Astronomy, Yale University, 52 Hillhouse, New Haven, CT 06511, USA}

\begin{abstract}

At present, there exists no consensus in the astronomical community regarding either the bulk composition or the formation mechanism for the interstellar object 1I/2017 U1 (`Oumuamua). With the goal of assessing the merits of the various scenarios that have been suggested to explain `Oumuamua's appearance and observed properties, we report a number of new analyses and provide an up-to-date review of the current hypotheses. We consider the interpretations that can reconcile `Oumuamua's observed non-Keplerian trajectory with the non-detection of traditional cometary volatiles. We examine the ability of these proposed formation pathways to populate the galaxy with sufficient interstellar objects such that the detection of `Oumuamua by Pan-STARRS would be statistically-favored. We consider two exotic ices, hydrogen and nitrogen, showing that the frigid temperature requirement for the former and the necessary formation efficiency of the latter pose serious difficulties for these interpretations. Via order-of-magnitude arguments and hydrodynamical cratering simulations, we show that impacts on extrasolar Kuiper Belt analogues are not expected to generate \ntwo{} ice fragments as large as `Oumuamua. In addition, we discuss observational tests to confirm the presence of these ices in future interstellar objects. Next, we examine the explanations that attribute `Oumuamua's properties to other compositions: ultra-porous dust aggregates and thin membranes powered by solar radiation pressure, among others. While none of these hypotheses are perfectly satisfactory, we make predictions that will be testable by the Vera Rubin Observatory to resolve the tension introduced by `Oumuamua.

\end{abstract}

% Add official keywords
\keywords{}

\section{Introduction} \label{sec:intro}

1I/2017 U1 (`Oumuamua), the first-detected decameter-sized object of interstellar origin within the Solar System \citep{Meech2017}, has evaded easy explanation as a direct analogue to the known populations of minor bodies. `Oumuamua's lightcurve varied quasi-periodically by approximately three magnitudes over a timescale of several hours \citep{Belton2018}, indicating an elongated geometry \citep{bannister2017col,Knight2017,Jewitt2017,Bolin2017} and the possibility of chaotic tumbling \citep{drahus2018tumbling,fraser2018tumbling}. Further analysis by \cite{mashchenko2019modelling} demonstrated that an oblate 6:6:1 aspect ratio was the most likely geometry. Immediate follow-up from ground-based instruments \citep{Meech2017} and ensuing space-based images from the \textit{Hubble Space Telescope} \citep{micheli2018non} and the \textit{Spitzer Space Telescope} \citep{Trilling2018} placed severe upper bounds on micron-sized dust and outgassing activity from CO and CO$_2$. Nonetheless, astrometric data indicated that `Oumuamua experienced non-gravitational acceleration with magnitude of order $10^{-3}$ of the Sun's gravity \citep{micheli2018non}. Thus, `Oumuamua's deviation from a ballistic trajectory and its red color \citep{bannister2017col,Jewitt2017,Fitzsimmons2017,Masiero2017} corroborate a cometary explanation, but the non-detection of the expected volatiles and the object's inferred elongation are problematic for this hypothesis.

Moreover, `Oumuamua's low inbound velocity with respect to the local standard of rest (LSR) implies that the object had experienced a dearth of stellar encounters during its interstellar journey \citep{hallatt2020dynamics}. Extrasolar systems typically have velocities close to the LSR upon formation that increase over time due to these scattering events. If `Oumuamua's age were similar to that of a typical Milky Way thin disk star, then its velocity with respect to the LSR should have been a few tens of km/s. Instead, `Ouamuamua's remarkably small incoming velocity indicated that it may be less than $100\,\text{Myr}$ old \citep{mamajek2017,gaidos2018and, feng2018oumuamua, hallatt2020dynamics}.

Oort Cloud objects located outside of the heliopause have persisted for the lifetime of the solar system. Therefore, analagous objects ejected from protoplanetary disks into interstellar space should persist for much longer than the $100\,\text{Myr}$ kinematically-estimated age of `Oumuamua. Explanations of this ISO's properties that invoke a solar system analog -- either a cometary or asteroidal composition -- would require `Oumuamua to have coincidentally been among the most recently-formed of its class. While the solar system's current position near the galactic midplane would result in a bias towards encountering younger ISOs in the local interstellar medium (ISM) \citep{hsieh2021colcar}, `Oumuamua's inbound velocity would still make it a genuine outlier.

In addition to these confounding physical and dynamical properties, the identification by Pan-STARRS of `Oumuamua, an indisputable interstellar object (ISO) passing within 60 lunar distances of the Earth, strongly suggests that a large galactic reservoir of similar bodies exists \citep{Do2018,trilling2017implications,Laughlin2017,Jewitt2017}. Therefore, a compelling explanation of `Oumuamua should permit the population of this reservoir under reasonable astrophysical circumstances.

While estimates of the ISO population and of the probability of detecting one such object varied widely before `Oumuamua \citep{sekanina1976probability, mcglynn1989nondetection,francis2005demographics, moro2009will, cook2016realistic}, the idea of using Pan-STARRS to bound their number density was introduced by \cite{jewitt2003project}. Indeed, just a few months prior to the encounter with \om, \cite{engelhardt2017observational} drew on the lack of detections of interlopers by Pan-STARRS, the Mt. Lemmon Survey, and the Catalina Sky Survey to determine a 90\% confidence upper limit on the population density of ISOs larger than $1\,\text{km}$ and without a brightness-boosting coma as $n_{\text{iso}} \sim 2.4\times10^{-2}\,\text{AU}^{-3}$. Thus, the ensuing discovery of `Oumuamua -- a smaller object passing closer to the Earth and found sooner than expected -- forced the reconsideration of both the number and the nature of ISOs.

While a reliable census of small interstellar bodies must wait until the forthcoming Vera Rubin Observatory (VRO) \citep{cook2016realistic, ivezic2019lsst}, a detailed analysis of the Pan-STARRS survey volume by \cite{Do2018} concluded a most-likely number density of $n_{\text{iso}} \sim 0.2\,\text{AU}^{-3}$. While the study assumes a perfect detection efficiency for `Oumuamua-like objects in the spatial region searched, accounting for instrumental imperfection would not substantially increase the estimate. \cite{portegies2018origin} reported a similar value of $n_{\text{iso}} \sim 0.08\,\text{AU}^{-3}$, with 95\% confidence interval spanning $0.004\,\text{AU}^{-3}$ to $0.24\,\text{AU}^{-3}$, using numerical simulations of protostellar disk ejecta.

In this study, we examine the ability of the leading exotic ISO hypotheses to generate a galactic population of `Oumuamua-like objects. While 2I/Borisov is also definitively of interstellar origin, its character was quickly determined to be analogous to Solar System comets \citep{Jewitt2019b}. Because `Oumuamua's anomalous properties are unique compared to the characteristics of the known minor bodies, including Borisov, we take it to be representative of its own distinct class. We do not consider distributions in composition, size, or inbound kinematics, and instead make the simplifying assumption that the reservoir of `Oumuamua-like ISOs consists of identical objects. Pan-STARRS has operated for a few additional years at higher efficiency without discovering another `Oumuamua-like ISO since the publication of \cite{Do2018}, so we set our fiducial value to $n_{\text{iso}} \sim 0.1\,\text{AU}^{-3}$ \citep{eubanks2021interstellar}. Although the statistics of one object are understandably uncertain, we  determine sensible confidence intervals in Table \ref{tab:uncertainty} via the classic method of \cite{garwood1936fiducial}.

\begin{table}[]
    \centering
    \begin{tabular}{c|c|c}
        Confidence & Lower Bound & Upper Bound \\
        \hline
        1$\sigma$ & $2\times10^{-2}\,\text{au}^{-3}$ & $0.3\,\text{au}^{-3}$ \\
        2$\sigma$ & $2\times10^{-3}\,\text{au}^{-3}$ & $0.6\,\text{au}^{-3}$ \\
        3$\sigma$ & $2\times10^{-4}\,\text{au}^{-3}$ & $0.8\,\text{au}^{-3}$ \\
        4$\sigma$ & $3\times10^{-6}\,\text{au}^{-3}$ & $1\,\text{au}^{-3}$ \\
    \end{tabular}
    \caption{Confidence intervals on the number density of \om-like objects in the solar neighborhood, assuming a most likely value of $n_{\text{iso}} \sim 0.1\,\text{au}^{-3}$ \cite{Do2018}.}
    \label{tab:uncertainty}
\end{table}

First, we discuss a general formalism to quantify the galactic population of \om-like objects in Section \ref{sec:oom}. Because each exotic hypothesis draws on different proposed astrophysical phenomena, it is necessary to develop a basis for direct comparison. With this framework, we analyze hypotheses relying on two exotic ices: \htwo{} in Section \ref{sec:h2}, as proposed by \cite{fuglistaler2018solid} and further examined by \cite{Seligman2020} and \cite{levine2021assessing}, and \ntwo{} in Section \ref{sec:n2}, as investigated by \cite{jackson20211i} and \cite{desch20211i}. In Section \ref{sec:observation} we discuss observations that may confirm the presence of either of these volatiles in future `Oumuamua-like objects.

Next, we consider other interpretations for \om{} in Section \ref{sec:nonice}. While we focus on ones involving propulsion via solar radiation pressure \citep{bialy2018could, moro2019could, luu2020oumuamua} to provide contrast with exotic ice outgassing, we also discuss  the possibility of CO ice as a solar system analogue \citep{seligman2021CO} and the tidal fragmentation explanation \citep{raymond2018implications,zhang2020tidal}, among others. Finally, in Section \ref{sec:discussion}, we discuss the implications of the detection of 2I/Borisov on these analyses and set broad expectations for VRO if any of these hypotheses for `Oumuamua are correct.

\section{Population of `Oumuamua-Like ISOs} \label{sec:oom}

We propose an order-of-magnitude model to estimate the galactic population of `Oumuamua-like objects from a given formation pathway. From this foundation, we discuss the required protoplanetary disk productivity to account for the inferred ISO reservoir.

\subsection{Formalism for Galactic Number Density}

The galactic number density, $n_{\text{iso}}$, of a given transient interstellar object class can be expressed as,

\begin{equation} \label{eq:drake}
    n_{\mathrm{iso}} = R_{\mathrm{env}} N_{\mathrm{iso}} t_{\text{iso}}\, V_{\mathrm{g}}^{-1}\,,
\end{equation}

\noindent where $R_{\text{env}}$ is the formation rate of the origin environment -- starless molecular cloud cores for \htwo{} icebergs and extrasolar systems for most other interpretations -- $N_{\text{iso}}$ is the average number of \om-like bodies ejected into the ISM from each birth system, $t_{\text{iso}}$ is the lifetime of the ISOs, and $V_{\text{g}} \sim 10^{67}\,\text{cm}^{3}$ is the volume of the galaxy. If $t_{\text{iso}}$ is greater than the Hubble time, then we substitute $N_{\text{env}}$, the number of ISO-spawning regions that have ever existed in the Milky Way, for $R_{\text{env}} t_{\text{iso}}$. In this situation, we would have,

\begin{equation} \label{eq:drakelonglasting}
    n_{\text{iso}} = N_{\text{env}}N_{\text{iso}}V_{\text{g}}^{-1}\,,
\end{equation}
for long-lasting ISOs.

Among the parameters in Equations \ref{eq:drake} and \ref{eq:drakelonglasting}, $R_{\text{env}}$ is the best-constrained for all birthplaces of interest in the Milky Way. Furthermore, the lifetime $t_{\text{iso}}$ can be reasonably-estimated via the material properties of the proposed compositions and knowledge of conditions in the local ISM. In contrast, $N_{\text{iso}}$ remains relatively unknown. Therefore, much of this paper is dedicated to determining and reconciling this value with each hypothesis.

\subsection{Required Production from Protoplanetary Disks}

Except for the \htwo{} ice interpretation, each explanation of `Oumuamua relies upon ISOs that  formed in extrasolar systems. Therefore, each of these protoplanetary ``disk product" hypotheses demands the production of a similar number of objects from the typical star, adjusted by the lifetime of the characteristic material.

From our fiducial $n_{\text{iso}}$ and $V_{\text{g}}$, we find that the Milky Way should currently contain of order $3\times10^{26}$ unbound `Oumuamua-like objects. We assume that the galaxy has generated ISOs from protostellar disks over the last $10\,\text{Gyr}$ with an average star-formation rate of $\sim3\,\text{yr}^{-1}$ \citep{Shu1987} and find that each system must produce of order $N_{\text{iso}} \sim 10^{16}$ objects. If these bodies have effective spherical radius $r_{\text{iso}}$ and bulk density $\rho_{\text{iso}}$, which we fiducially take to be $100\,\text{m}$ and $1\,\text{g}\,\text{cm}^{-3}$, respectively, then we estimate the total mass $M_{\text{tot}}$ that must be ejected per star to be approximately

\begin{equation} \label{eq:diskscaling}
    M_{\text{tot}} \approx 7\text{\mearth}\,
    \Bigg(\frac{r_{\text{iso}}}{100\,\text{m}}\Bigg)^{3}
    \Bigg(\frac{\rho_{\text{iso}}}{1\,\text{g}\,\text{cm}^{-3}}\Bigg)\,.
\end{equation}

\noindent This value is generally concordant with models of protoplanetary disk evolution. It is important to note that due to `Oumuamua's poorly-constrained albedo, physical size, and density, this required production could vary by orders of magnitude. Here, the fiducial size which we have adopted corresponds to a mass of order $10^{12}\,\text{g}$ for an individual interstellar object.

In addition to its young inferred age, `Oumuamua's inbound kinematics were temporally and spatially correlated with the Columba and Carina (COL/CAR) young stellar associations \citep{mamajek2017kinematics,gaidos2018and, hallatt2020dynamics, hsieh2021colcar}. That is, the ISO was in these regions while they were forming new stars, so it is suggestive that `Oumuamua may have formed there as well. Therefore, we consider the possibility of detecting a small body analogue specifically from COL/CAR in the Pan-STARRS survey. Because these associations are younger than the galactic rotation period, the ISOs from these regions should only fill a volume comparable to that of the stars that comprise these moving groups. For this situation, we can modify Equation \ref{eq:drakelonglasting} by substituting the present volume of the COL/CAR stellar associations, $V_{\text{local}} \sim 10^{6}\,\text{pc}^{3}$ \citep{hsieh2021colcar}, for $V_{\text{g}}$, and only consider the number density in the Solar neighborhood.

While the number of counted members in stellar associations varies based on the classification methodology, we adopt the value of $N_{\text{env}} = 100$ stars between COL/CAR as an order of magnitude approximation to the data presented in Table 3 of  \citet{hallatt2020dynamics}. We assume that each protoplanetary disk ejects the fiducial $7\,\text{\mearth}$ of primordial debris in the form of $10^{12}\,\text{g}$ interstellar comets, as found in Equation \ref{eq:diskscaling}. Spreading these objects uniformly over $V_{\text{local}}$ provides a number density of $5\times10^{-4}\,\text{AU}^{-3}$ for ISOs that are conventional solar system analogues. Detecting a planetesimal from COL/CAR with Pan-STARRS would be between a 2-3$\sigma$ event, according to our calculations in Table \ref{tab:uncertainty}.

\section{Hydrogen Icebergs} \label{sec:h2}

Solid hydrogen was first discussed as a possible composition for interstellar interlopers by \cite{fuglistaler2018solid}. \cite{Seligman2020} advanced this hypothesis by demonstrating the consistency of \om's observed non-gravitational acceleration and non-detection of a coma with that of an object covered by \htwo{} ice on approximately 6\% of its surface. Furthermore, the volatility of this ice in the ISM dictates that these exotic icebergs should be recently-formed while still being sufficiently-eroded to develop an extreme aspect ratio \citep{domokos2017explaining}. Thus, the \htwo{} interpretation explains `Oumuamua as the product of COL/CAR and unassociated with any successfully-formed star: a ``cloud product."

\begin{figure*}
\begin{center}
    \includegraphics[scale=0.6,angle=0]{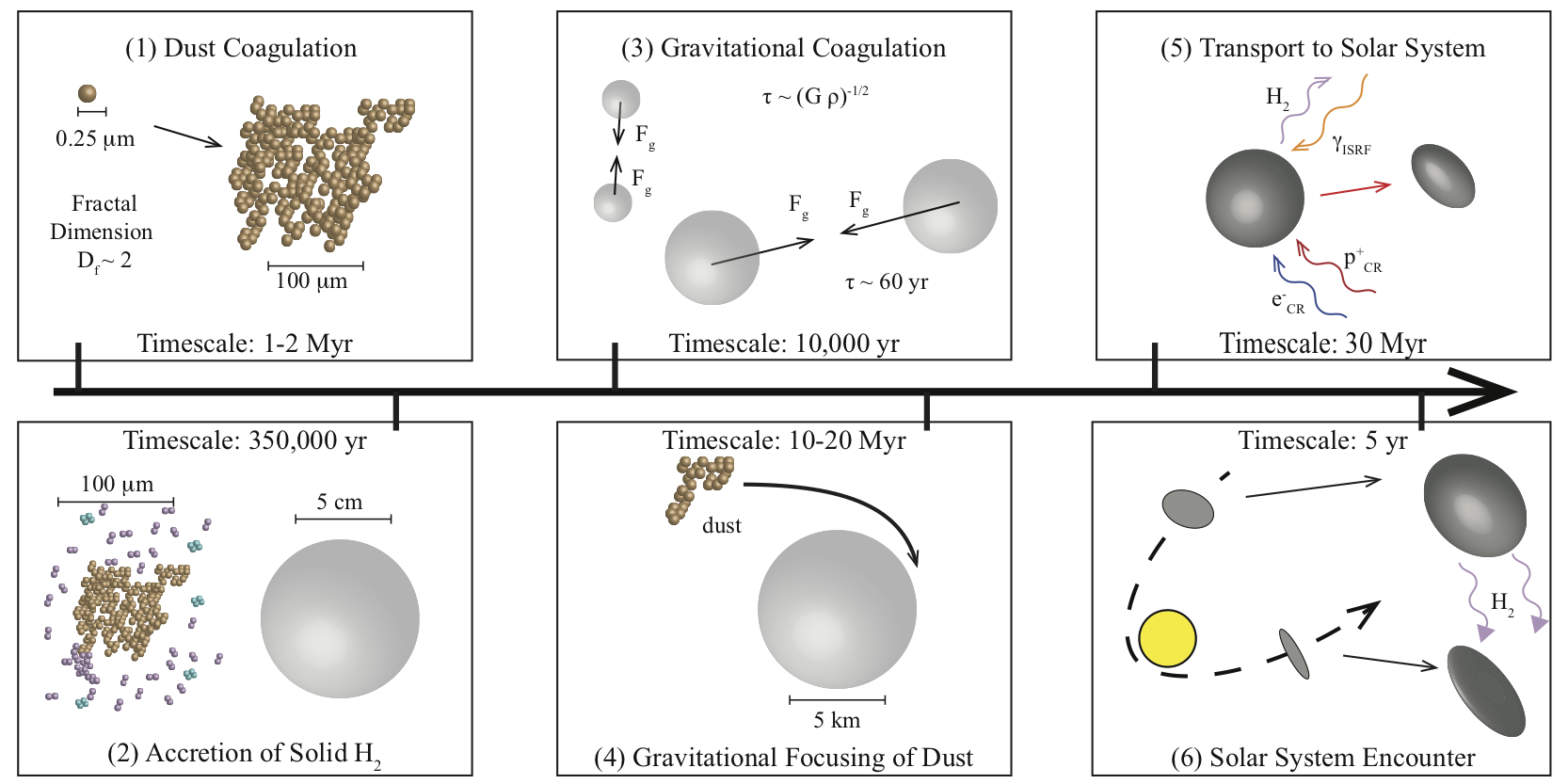}
    \caption{Schematic illustration of the formation mechanism for hypothetical `Oumuamua-like progenitors with a solid \htwo{} component}, which was proposed in  \cite{levine2021assessing}. Panel (1) shows the coagulated, porous, non-ideal dust required for the grain temperature to approach the $2.8\,\text{K}$ needed for hydrogen deposition. Then, panel (2) shows the accretion of \htwo{} and micron-sized dust until the growing ``hailstones" decouple from the starless core's turbulence and aggregate into ``snowballs" via gravitational free-fall (panel 3). In order to retain some hydrogen ice for longer timescales as they travel through the ISM, the kilometer-sized progenitors must aggregate additional micron-sized dust to bring their metallicity above the cosmic abundance (panel 4). As the icebergs are transported through the galaxy, they lose \htwo{} via mostly thermal sublimation and some cosmic ray bombardment (panel 5). Finally, the Solar encounter dramatically increases `Oumuamua's aspect ratio (panel 6), via the \htwo{} outgassing that propels the non-gravitational acceleration.
    \label{fig:h2_schematic}
\end{center}
\end{figure*}

\subsection{Generation of Interstellar Hydrogen Icebergs}

\cite{levine2021assessing} considered theoretical constraints on forming ISOs composed of solid hydrogen   at size regimes ranging from micron-sized dust to centimeter-sized hailstones to the kilometer-scale aggregates that would be required to form `Oumuamua-like objects. Figure \ref{fig:h2_schematic} summarizes the examined steps in a hypothetical formation pathway from that study to form \htwo{} objects. Because of the $3\,\text{K}$ temperature requirement for \htwo{} deposition, \cite{levine2021assessing} considered the starless cores of giant molecular clouds (GMCs) as formation sites. These frigid temperatures have not been observed in starless cores specifically, although CO(J=1-0) excitation temperatures of $1\,\text{K}$ have been recorded in the outflows in pre-planetary nebulae \citep{sahai2017coldest}. While this observation did not imply that quiescent molecular cloud cores can reach these temperatures, it  suggested that adiabatic expansion may cool regions below the temperature of the cosmic microwave background (CMB). \cite{levine2021assessing} found that \htwo{} could only freeze onto coagulated dust grains in pockets of adiabatic expansion that are well-shielded from electromagnetic radiation.

Nonetheless, \cite{levine2021assessing} found that the turbulence characteristic to starless cores may be conducive to forming kilometer-scale \htwo{} objects that, when weathered, could have properties resembling the size, shape, and composition of \om. If these hydrogen icebergs can grow to sizes such that they attract additional dust via gravitational focusing, they could retain enough ice \citep{walker2013snowflake} after a journey through the ISM to match `Oumuamua's non-gravitational acceleration during the Solar encounter. \cite{hoang2020destruction} quantified \htwo{} destruction via drag heating, but this process did not necessarily apply to `Oumuamua because of its small velocity versus the ambient interstellar matter. Since starless cores are probably magnetically supported \citep{Shu1987, kong2016deuterium}, it is possible that dust plasma instabilities, such as magnetic resonant drag instabilities \citep{Seligman2019_RDI,Hopkins2020}, also play a role in the formation of these hydrogen icebergs.

\subsection{Occurrence of Solid Hydrogen Interstellar Objects}

The formidable temperature barrier implies that $N_{\text{iso}}$ in Equation \ref{eq:drake} may be zero for interstellar \htwo{} icebergs. If these ISOs do exist, however, the large amount of hydrogen gas in GMCs would suggest that `Oumuamua-like cloud products could form via an inefficient formation process compared to the requirements for disk products. Nevertheless, the rapid sublimation in the ISM requires that these icebergs would have larger masses upon formation than the objects in other hypotheses. This would increase the implied mass of the galactic population and potentially contradict the limits from the mass budget, despite the larger overall hydrogen reservoir versus that of other substances. `Oumuamua's correlation with COL/CAR, combined with the sub-$100\,\text{Myr}$ survival timescale of interstellar solid hydrogen, makes it improbable that an ISO with substantial \htwo{} passing by Earth would have originated from anywhere other than these associations.

\cite{levine2021assessing} estimated the efficiency of building \htwo{} ice objects in the overall GMC complexes to compare with the star-formation efficiency, and \cite{hsieh2021colcar} discussed forming objects of `Oumuamua's present size in COL/CAR. Here, we discuss creating the required kilometer-scale progenitors in the failed cores. We assume, perhaps conservatively, that there were a similar number of star-forming and starless cores \citep{vazquez2005lifetimes}, the latter of which could produce \htwo{} rich ISOs. Taking the characteristic lifetime of `Oumuamua-like objects to be longer than the age of these associations, we then set $N_{\text{env}} = 100$ and apply Equation \ref{eq:drakelonglasting}.

To match the inferred ISO number density with the kilometer-scale progenitors from \cite{levine2021assessing}, each starless core must generate of order $10^{19}$ ISOs. Because rapid \htwo{} sublimation demands that these cosmic icebergs must be created with mass of order $10^{15}\,\text{g}$ to resemble `Oumuamua when they reach the Solar System, this productivity would result in each core forming $4\,\text{M}_{\odot}$ of \htwo{} ice. This value is too high to reasonably be formed in small pockets of adiabatic expansion, although the 95\% confidence interval's lower bound would only require that $0.1\,\text{M}_{\odot}$ of \htwo{} ice objects are born in each core.

The \htwo{} interpretation predicts the presence of both actively-outgassing ISOs and their remnants. Because \htwo{} ice would be required to drive the non-gravitational acceleration, `Oumuamua would necessarily be a member of the former population in this hypothesis. Moreover, the objects rich in \htwo{} would necessarily be younger and found closer to their birthplaces since sublimation removes the volatile, exotic ice on timescales faster than the galactic rotation period. In contrast, the long-lived remnant population could spread uniformly in the galaxy. The composition of the remnants would depend both on their origin environment and on the interstellar conditions though which they pass. Without their original \htwo{} ice endowment, however, their compositions would most likely be dominated by a combination of refractory and icy material acquired in the molecular cloud of origin.

If `Oumuamua were an exotic \htwo{} iceberg, the fact that it would have still contained significant \htwo{} implies that the local concentration of active objects is higher than the average galactic number density of their inactive descendants. Furthermore, the stringent thermodynamic requirement for hydrogen deposition and increasing CMB temperature with redshift means that \htwo{} ISOs would be recent phenomena in galactic history \citep{white1996dark}. For example, \cite{levine2021assessing} found that \htwo{} deposition would be infeasible at reasonable starless core pressure for redshift $z \gtrsim 0.1$. We assume that the remnants without \htwo{} survive indefinitely in the ISM. Although they could experience breakup during stellar encounters, as seen in Solar System comets, the rate of  close encounters for any given ISO are exceedingly low. Thus, the ratio of active to inactive cloud product ISOs would constrain the lookback time at which \htwo{} ice became possible in the Milky Way.

Assuming the same productivity, $N_{\text{iso}}$, per starless core for as long as \htwo{} deposition has been feasible, we compare the number density of `Oumuamua-like objects from COL/CAR to that of their remnants to find,

\begin{equation}
    f\,\bigg(\frac{N_{\text{env}}}{V_{\text{local}}}\bigg) =\bigg( \frac{R_{\text{core}}t_{\text{look}}}{V_{\text{gal}}}\bigg)\,.
\end{equation}

%\begin{equation}
%    N_{\text{core}} = f\, \bigg(\frac{ N_{\text{env}} %V_{\text{gal}}}{V_{\text{local}}}\bigg)\,,
%\end{equation}

\noindent Here, $f$ is the ratio between these two number densities and $N_{\text{env}} = 100$ is the assumed number of starless cores for COL/CAR. Additionally, $R_{\text{core}}t_{\text{look}}$ is the rate of starless core formation multiplied by the lookback time at which \htwo{} iceberg formation became feasible, giving the number of systems that would have created `Oumuamua-like objects over Galactic history. We have eliminated $N_{\text{iso}}$, implicitly assuming that it would be the same between COL/CAR and any other starless cores that form hydrogen ISOs. If the rate of failed cores is similar to star-producing ones, then the Milky Way can generate an overall ISO number density equal to the local one from COL/CAR ($f = 1$) in $t_{\text{look}} \sim 10\,\text{Myr}$.

This rapid formation of a Galactic population is shorter than the age of the COL/CAR associations. If the \htwo{} hypothesis is correct, then it would be statistically unlikely to observe an active iceberg instead of a remnant from the broader Galaxy. While this tension could be resolved if the failed cores in these specific moving groups became unusually cold, this degree of fine-tuning is not supportive of the \htwo{} hypothesis. Explanations requiring that the Solar System inhabit a special region are generally disfavored, and the hydrogen ice interpretation would demand this conclusion to be compatible with the inferred ISO population.

\section{Nitrogen Icebergs} \label{sec:n2}

While focusing on \htwo, \cite{Seligman2020} showed that \ntwo{} outgassing would also be compatible with `Oumuamua's non-Keplerian trajectory. This idea was furthered when \cite{jackson20211i} suggested that these exotic icebergs could form as collisional shards from extrasolar Pluto analogues. In this case, `Oumuamua would be nearly pure nitrogen, with higher albedo ($g \sim 0.65$) and smaller size ($45\,\text{m}$) than assumed for other hypotheses. The geochemical and dynamical requirements to build these objects and transport them through the galaxy were examined by \cite{desch20211i}.

This interpretation is appealing because it requires only substances that exist on Kuiper Belt Objects (KBOs). However, far less of the Milky Way's mass is in nitrogen ice compared to molecular hydrogen. Thus, a greater proportion of the \ntwo{} would need to be converted to `Oumuamua-like ISOs compared to the fraction of \htwo{} that would need to freeze. For example, \cite{siraj2021mass} found that an unreasonably large galactic production of KBO analogues would be required to produce the inferred population of \om-like objects through the formation model of \cite{desch20211i}. Moreover, the reservoir of typical cometary volatiles in protoplanetary disks is also orders-of-magnitude larger than that of \ntwo{} ice. 

Here, we take a different and complementary approach to reassess bounds on the galactic reservoir of \ntwo{} collisional fragments by discussing the generation of these objects on extrasolar Pluto-sized KBO analogues. Specifically, we focus on the impacts required to produce the ejecta. We find that if `Oumuamua were one such shard, then its appearance in the solar system and detection by Pan-STARRS would have been statistically-disfavored. Thus, VRO should not expect to find any similar objects in the future.

\subsection{Galactic Reservoir of Nitrogen Fragments}

\cite{desch20211i} estimate that our Solar System ejected $N_{\text{iso}} \sim 2\times10^{14}$ nitrogen ice collisional fragments in their Table 1, and then  generalize this calculation to estimate the  population of fragments in the Milky Way. We compactly re-express this calculation as

\begin{equation} \label{eq:n2population}
    N_{\mathrm{iso}} = \frac{M_{\mathrm{kb}}f_{\mathrm{big}}f_{\mathrm{N}}f_{\text{\ntwo}}f_{\mathrm{surf}}f_{\mathrm{stick}}f_{\mathrm{coll}}f_{\mathrm{ej}}f_{\mathrm{ss}}}{M_{\mathrm{iso}}}\,,
\end{equation}

where the terms are defined as follows:

\begin{itemize}
    \item $M_{\text{kb}}$: the mass of a typical extrasolar Kuiper Belt region.
    \item $f_{\text{big}}$: the mass fraction of the Kuiper Belt residing in objects large enough to purify nitrogen via hydrothermal chemistry.
    \item $f_{\text{N}}$: the mass fraction of nitrogen in these large KBO analogues.
    \item $f_{\text{\ntwo}}$: the mass fraction of nitrogen on these objects converted to \ntwo.
    \item $f_{\text{surf}}$: the mass fraction of molecular nitrogen that outgasses and freezes in a surface ice layer.
    \item $f_{\text{stick}}$: the mass fraction of this surface \ntwo{} ice layer that does not sublimate before the dynamical instability.
    \item $f_{\text{coll}}$: the mass fraction of the remaining surface nitrogen ice that is excavated from the KBOs via collisions.
    \item $f_{\text{ej}}$: the fraction of collisional shards from impacts that are ejected from the Solar System.
    \item $f_{\text{ss}}$: the stellar mass-weighted fraction of extrasolar systems that satisfy the conditions to form nitrogen fragments in this manner.
    \item $M_{\text{iso}}$: the mass of the \ntwo{} collisional fragments upon ejection.
\end{itemize}

In their study, \cite{desch20211i} assumed that a Kuiper Belt of mass $M_{\text{kb}} \sim 35\,\text{\mearth}$ would contain approximately $6\,\text{\mearth}$ of material in objects large enough to support thick layers of surface \ntwo{} ice ($f_{\text{big}} \sim 0.8$). Given a mass fraction of nitrogen $f_{\text{N}} \sim 0.05$, the perfect efficiency of geochemical processes in transporting pure \ntwo{} to the surface ($f_{\text{\ntwo}} = f_{\text{surf}} = 1$), and the expectation that these KBO analogues would retain all of this surface ice ($f_{\text{stick}} = 1$), then about $0.3\,\text{\mearth}$ of \ntwo{} ice would be exposed before impacts are capable of ejecting \ntwo{} fragments.

During the ensuing and required dynamical instability of the giant planets, \cite{desch20211i} proposed that $0.043\,\text{\mearth}$ of \ntwo{} collisional fragments are generated and ejected into interplanetary space ($f_{\text{coll}} \sim 0.15$). Of this mass, the study proposes that $0.008\,\text{\mearth}$ survives to be ejected into interstellar space ($f_{\text{ej}} \sim 0.19$) as shards with mass $M_{\text{iso}} \sim 2\,\times 10^{11}\,\text{g}$. In addition, \cite{desch20211i} suggested that only GKM stars can produce these fragments, so $f_{\text{ss}} \sim 0.5$. Therefore, the resulting number of ISOs would be of order $N_{\text{iso}} \sim 1\times10^{14}$ per exoplanetary system for galaxy-wide production. Because N$_2$ icebergs survive longer than the rotation period of the Milky Way, the spatial distribution of these ISOs should be nearly isotropic.

Considering a lifetime for \ntwo{} icebergs of $t_{\text{iso}} \sim 2\,\text{Gyr}$, the optimistic upper bound on `Oumuamua's age from \cite{jackson20211i}, with our assumed values for the star-formation rate $R_{\text{env}}$ and $V_{\text{g}}$, we find $n_{\text{iso}} \sim 2\times 10^{-4}\,\text{AU}^{-3}$. This value is within a factor of a few of the one provided in Table 1 of \cite{desch20211i} specifically for nitrogen ice fragments. Nonetheless, \cite{desch20211i} claim that the Galactic reservoir of nitrogen shards corroborates the $2\sigma$ lower bound from \cite{portegies2018origin}. However, the number density applied for this comparison, of order $10^{-3}\,\text{AU}^{-3}$, corresponds to the total population of shards and not just those of nitrogen ice. Considering only this subset revises $n_{\text{iso}}$ down by a factor of 10 and places the observation of one such body by Pan-STARRS around the lower bound of the $3\sigma$ confidence interval. Thus, even with the assumed widespread geochemical differentiation of extrasolar KBOs and ubiquity of dynamical instabilities, `Oumuamua's appearance would still be quite unexpected.

\subsection{Upper Bound on the Nitrogen Iceberg Population}

Given the unsatisfactorily low $n_{\text{iso}}$ if `Oumuamua-like \ntwo{} fragments are formed via the mechanism proposed in \citet{desch20211i}, we consider the extreme case for  the maximum theoretical production from exo-KBOs. We set each fraction, $f$, in Equation \ref{eq:n2population} to unity except for $f_{\text{N}}$, which we leave as $0.05$ in-line with expectations from cosmic abundances \citep{draine2011book}. From this estimate, we find $N_{\text{iso}} \sim 5\times10^{16}$ and $n_{\text{iso}} \sim 0.1\,\text{au}^{-3}$. Essentially, reconciling the $n_{\text{iso}}$ estimate from \cite{Do2018} requires that all extrasolar Kuiper Belts convert the entirety of their cosmic abundance of nitrogen into `Oumuamua-sized ISOs. Immediately, the widespread \ntwo{} ice on present-day Pluto and Triton rules out this scenario for our Solar System.

In addition, we  determine the fraction of nitrogen in the minimum mass Solar nebula (MMSN) that must be converted to \ntwo{} icebergs to satisfy the inferred $n_{\text{iso}}$. \cite{siraj2021mass} also examined constraints based on the MMSN, but investigated the mass budget required to build the KBO progenitors to \ntwo{} fragments. Here, we consider the more ambitious upper bound of converting all available nitrogen to `Oumuamua-like shards.

In both classic \citep{weidenschilling1977distribution} and more recent MMSN \citep{desch2007mass} models, the protoplanetary disk mass is never above $0.1\,\text{\msun}$. Assuming that the entire cosmic abundance of nitrogen in this upper limit is converted to icebergs, we find the resulting number density would be of order $1\,\text{AU}^{-3}$. For this exercise to reflect the true occurrence of `Oumuamua-like objects, it would require condensation and purification of \ntwo{} far inside its ice line through unknown processes. If `Oumuamua is representative of the most likely value for the galactic population of these \ntwo{} icebergs, then approximately 10\% of all nitrogen in protoplanetary disks must be incorporated in  these ISOs. 

%In the Milky Way, the \cite{Do2018} number density would correspond to 0.1\% of the interstellar nitrogen budget being locked-up in `Oumuamua-like objects.

\subsection{Formation of Icy Fragments from Impacts}

The `Oumuamua-like ISO formation pathway from \cite{desch20211i} requires impacts on bodies with purified surface N$_2$ ice: analogues of KBOs at least the size of Gonggong ($\sim 600$ km). Impact cratering is thoroughly described by \citet{Melosh1989}, including relationships between ejecta and projectile dimensions. While studies of the bulk properties of N$_2$ ice are limited, they are sufficient to acquire order-of-magnitude estimates of the impactor, crater, and ejecta characteristics. 

Lightly shocked surface material is ejected in a process known as spallation \citep{Melosh1984}. The interference between the compressional shock wave delivered by the impact and the tensional rarefaction wave that reflects off the free surface reduces the maximum pressure experienced by the material. However, the surface layer is accelerated upward by the pressure gradient from both waves, which enables some material to reach the escape velocity. Spall layers are the source of the large ejecta blocks, and the remainder of the ejecta is produced by the excavation flow of highly-shocked material. \citet{desch20211i} attribute ejected material to the excavation flow; however,  excavation velocities are at most between $1/3$ and $1/5$ of the peak particle velocity \citep{Melosh1989}, which itself is upper-bounded by half the impact velocity, $U/2$. Therefore, it might be responsible for some secondary cratering effects, but is unlikely to be responsible for fragments that reach the escape velocity of the Pluto or Gonggong analogues. Implicit in the analysis of \cite{desch20211i} is that the entirety of the crater is ejected; however, one should restrict the analysis to the spall layer, which separates at the depth where the tensional pressure of the rarefaction wave equals the target's tensile strength $\sigma_t$. 
Per \citet{Melosh1984}, peak particle velocity $v_p $ a distance $r$ from the impact site may be calculated as, 
\begin{equation}
    v_p = C_V \frac{U}{2}\,\Bigg(\frac{a}{r}\Bigg)^{1.87},
\end{equation}
where the projectile radius and velocity are denoted as $a$ and $U$, respectively, the exponent 1.87 is found empirically \citep{Perret1975}, and the coupling constant $C_V$ approximately accounts for differences between projectile and target density. The pressure from the shock near the surface is approximately $P \simeq \rho_t c_L v_P$, for target density and sound speed, $\rho_t$ and $c_L$. Evaluating the pressures induced by the shock and rarefaction waves, and performing a Taylor expansion to first order in depth $z$, gives the depth of spall layer $z_s$ as \citep{Melosh1984}

\begin{equation}
     z_s \simeq \bigg(\frac{\sigma_t}{P(r)}\bigg)\,\bigg(\frac{\beta c_L \mathcal{T}}{2d}\bigg)\,\bigg(\frac{r}{1 - 1.87{\beta c_L \mathcal{T}/r}}\bigg)\,,
\end{equation}

\noindent Here, $d$ is the equivalent depth of the explosion, $\mathcal{T} = a/U$, and $\beta$ is a factor of order unity related to the decay time of the stress pulse ($d/c_L = \beta\mathcal{T}$). Different projectile and target materials are incorporated as $d\simeq2a\sqrt{\rho_p/\rho_t}$. Substituting in the ejection speed $v_e \approx 2 v_p d /r$ \citep{Melosh1984}, and making a further approximation $\mathcal{T} c_L \ll \sqrt{r^2-d^2}$, one arrives at,

\begin{equation} \label{eqn:spall}
    z_s \simeq\, \Bigg(\frac{\sigma_t}{\rho_t c_L}\Bigg) \,\bigg( \frac{d}{v_e}\bigg).
\end{equation}

This depth is an upper limit to the thickness of spall layers. Flaws or other structural properties of the target may cause shallower layers to fragment. For ejection velocities exceeding $\sim 0.5$ km s$^{-1}$, material within and below the spall layer experiences further fragmentation \citep{Grady1980, Grady1987, Melosh1989}, a trend reproduced in numerical simulations \citep{Melosh1992}. The mean size of Grady-Kipp fragments, $ l_{GK}$, may be estimated \citep{Melosh1989} through,
\begin{equation} \label{eqn:gk}
    l_{GK} = \,\bigg(\frac{\sigma_t}{\rho_t v_{e}^{2/3}\,U^{4/3}}\bigg)\,d.
\end{equation} 
Comparisons carried out by \citet{Artemieva2004} and \citet{McEwen2005} showed that this approximation agrees (within a factor of $\sim 2$) with more sophisticated models when predicting the largest size of ejecta fragments. 

% `Oumuamua's size is not well constrained, but the effective diameter is likely between $98-440$ m \citep{Trilling2018}. 
We consider the plausible sizes of fragments originating from a KBO impact, given the bulk properties of solid N$_2$. The maximum compressive stress of N$_2$ ice before fracture is about 9 MPa at 5 K, correpsonding to brittle failure mode \citep{Yamashita2010}. At larger temperatures, the ice switches to a ductile failure mode, and can only withstand compression of $\lesssim 1$ MPa. The transition occurs at temperatures $\sim30-40$ K, depending on the strain rate. 
We adopt $\sigma_t \sim 1.4$ MPa as the tensile strength of N$_2$ ice, as measured at $40\,\text{K}$ \citep{Pederson1998}. The minimum ejection speed required by the N$_2$ hypothesis for `Oumuamua is the escape velocity of Pluto, $v_e \geq 1.2$ km s$^{-1}$. Escape of spall material necessitates an impact of at least twice the escape velocity $U \geq 2.4$ km s$^{-1}$, if the projectile and target are composed of identical materials. At Pluto's distance from the Sun, projectiles should mainly be KBOs with density ranging between $1-3\,\text{g}\,\text{cm}^{-3}$ for high ice and rock fraction, respectively \citep{Brown2012}. The mean encounter speed of a KBO with Pluto is 2.46 km s$^{-1}$ \citep{Delloro2013}, which amounts to an impact speed of $2.73$ km s$^{-1}$. This speed is only marginally above the threshold needed to eject spall material; many impacts will not eject any at all.

\citet{desch20211i} argued that, in the early Solar System, small KBOs were dynamically excited to speeds of $\sim 3.5$ \kms\ relative to large KBOs. For the following analysis, we adopt a nominal impact speed of $U = 3.7$ km s$^{-1}$ for excited KBOs where Pluto's escape velocity has been added in quadrature. We note that the analytic model \citep{Melosh1984} suggests that this assumption of more violent impacts than in today's Kuiper Belt is required to possibly eject material that could be a progenitor of `Oumuamua. Additionally, we take the density of N$_2$ ice as $\rho_t \approx 1$ g cm$^{-3}$ \citep{Trowbridge2016} and the longitudinal sound speed $c_L \approx 1.8$ km s$^{-1}$ \citep{Yamashita2010}.

A spall layer 100 m thick ejected from Pluto requires $d \approx 150$ km, using Equation~\ref{eqn:spall}. Assuming an unlikely projectile consisting of only rock, its diameter must be at least $\sim 90$ km. As discussed above, the spall thickness is an upper bound of the typical size of high-speed ejecta, owing to the elastic energy within the spall plate and subsequent Grady-Kipp fragmentation (Equation~\ref{eqn:gk}). The majority of spalled mass will be in the form of these smaller fragments. Even a rocky projectile still must be $\sim 250$ km wide in order for 100 m ice chucks to survive finer fragmentation. In the case of a pure ice projectile, the diameter must be $\sim 450$ km. 

\begin{figure*}
  \includegraphics[width=\textwidth]{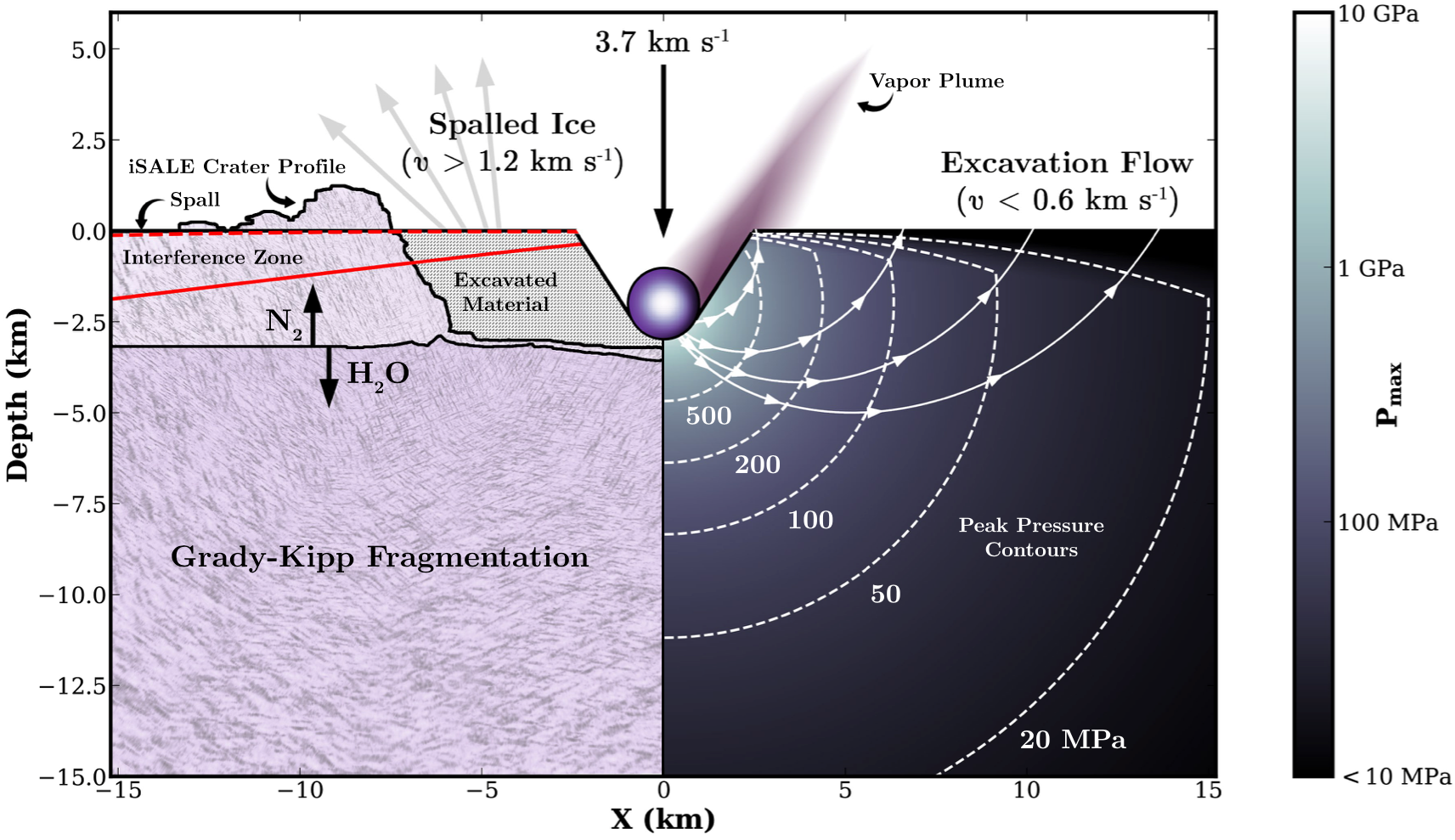}
  \caption{Schematic of a hypervelocity impact, adapted from \citet{Melosh1984}, to depict the case study of a KBO impact on Pluto. The 2 km diameter projectile comprises H$_2$O ice, and the target consists of H$_2$O underlying a 3 km N$_2$ ice surface layer. Contours on the right indicate the peak shock pressure experienced by the target according to the analytic wave-interference theory \citep{Melosh1984} (e.g. the contours exhibit a sharp, inwards turn near the surface due to interference between rarefaction and compression waves). Arrowed curves indicate the excavation flow, which can reach speeds up to $\sim 1/6$ of the impact speed. The spall zone, which represents high-speed and lightly-shocked material, is labeled on the left as a very thin surface layer. Additional features include a highly-shocked vapor plume that emanates from the point of impact, and Grady-Kipp fragments forming deep into the target. The figure includes results from the iSALE-2D simulation, including the crater profile and boundary between the two ices, both shown on the left.}
  \label{fig:impact}
\end{figure*}

Statistics for KBOs with diameter exceeding 100 km are given by \citet{Bierhaus2015}. They showed that an impact on Pluto by such an object is unlikely over the past 4 Gyr and that the largest expected impactor is about $60$ km. 
% The scenario is even less likely given `Oumuamua's proposed age under the N$_2$ model of 0.5 Gyr (Jackson \& Desch 2021). A very recent study \citep{Morbidelli2021} corroborates the rarity of such impact events.
\citet{Singer2019} interpreted craters imaged by {\it New Horizons} using Schmidt-Housen-Holsapple scaling relations \citep{Holsapple1993, Housen2011}. With the exception of Sputnik Basin (which may have formed by impact of a $150-275$ km diameter projectile, likely more than 4 Gyr ago), the remaining largest craters were formed by projectiles with diameters not exceeding $\sim 50$ km.

We consider whether large-scale ($>50$ km) impacts could be responsible for the N$_2$ ice population postulated by \citep{jackson20211i}. One may estimate the ratio of ejected spall mass, $m_{\rm ej}$ to projectile mass, $m_i$, \citep{Melosh1984, Melosh1985, Armstrong2002}, using,
\begin{equation}
\label{eqn:ratio}
   \frac{m_{\rm ej}}{m_i} = \frac{0.75P_{\rm max}}{\rho_t c_L U}\Bigg(\Big(\frac{U}{2v_{e}}\Big)^{5/3} - 1 \Bigg),
\end{equation}
where $P_{\rm max} \approx \beta \sigma_t$ is the maximum compressive pressure experienced by the ejecta ($\beta \sim 4$). For the nominal values reported above, and assuming an impact speed of $3.7$ km s$^{-1}$, we find $m_{\rm ej}/{m_i} \approx 7 \times 10^{-4}$. For example, an exceptionally large icy impactor of size $200$ km and density $1$ g cm$^{-3}$ would produce about $3 \times 10^{15}$ kg of spall, or $10^7$ fragments with mean radius $40,\text{m}$ (mass $\sim2.4 \times 10^8$ kg) determined from Equation \ref{eqn:gk}.
% This optimistic estimate for the number of `Oumuamua-sized fragments neglects subsequent weathering, which requires larger and even fewer ejecta. 
\citet{desch20211i} acknowledged that Pluto experienced of order unity (up to about 4) impacts of this scale; hence, there should be of order a few $\times 10^7$ spall fragments the size of `Oumuamua ejected from Pluto --- an estimate five orders of magnitude lower than the number that \citet{desch20211i} suggested are ejected into interstellar space. This striking difference arises mainly from three cratering-related processes:

\begin{enumerate}
    \item N$_2$ ice is relatively weak, and is easily fragmented into $<10\,\text{m}$ chunks during impacts exceeding $2-3$ \kms.
    \item At least $\gtrsim 2.4$ \kms\ impacts are needed to eject any lightly-shocked material, which in turn reduces the maximum size of ejected fragment.
    \item Only a fraction of a percent of the impactor mass is ejected from the KBO analogue.
\end{enumerate}

\noindent These considerations significantly revise-down the fragment production estimates of \citet{desch20211i}, which relied on the assumption that the entirety of the crater is ejected via excavation flow, and that $50\,\text{m}$ fragments are the most commonly-produced size.

\subsection{Hydrodynamical Impact Simulations}

In order to validate the estimates in the previous subsection, we simulate a KBO impact on Pluto with the iSALE-2D (Dellen) hydrodynamical code \citep{Collins2014}. The numerical setup is depicted in Figure~\ref{fig:impact}. The code is based on the SALE algorithm \citep{Amsden1980}, and incorporates a number of sophisticated material models \citep{Wunnemann2006, Melosh1992, Ivanov1997, Collins2004, Collins2011}. Our simulation employs a Tillotson Equation of State (EoS) for N$_2$. Specific heat and bulk modulus are obtained from \citet{Trowbridge2016} and \citet{Yamashita2010}, respectively. We use an initial density of 0.995 g cm$^{-3}$. The remaining EoS parameters are manually varied to fit to the Hugoniot profile used by \citet{Trowbridge2015}. While the internal energies of complete ($E_{CV}$) and incipient ($E_{IV}$) vaporization are not provided in the literature, we take $E_{CV}$ as one third of of the Tillotson $E_0$ parameter, and $E_{IV}$ as one fourth of $E_{CV}$, which are roughly the same ratios as in the case of water ice \citep{Melosh1989}. Our choices of $E_{IV}$ and $E_{CV}$ for N$_2$ are roughly an order of magnitude lower than that of water, reflecting N$_2$'s significantly higher volatility. Under the Tillotson EoS, iSALE-2D computes pressure from density and internal energy ($P = P(\rho, E)$) in three distinct regimes, each obeying a different functional form: compressed ($E < E_{IV}$), transition ($E_{CV} > E > E_{IV}$), and expanded ($E > E_{CV}$). Similar to \citet{Trowbridge2015}, we adopt a Lundborg strength model for N$_2$, using methane as a proxy for Simon melting parameters and shear strength \citep{Haynes1971}. An H$_2$O ice layer in the target is modeled with the ice Tilloston EoS included with iSALE-2D. For H$_2$O, we use the strength and damage models of \citet{Johnson2016}, with the same parameter values. 

It is emphasized that the simulations presented here are an approximation of real impact cratering processes. The simulations neglect potentially important parameters such as impact angle, the porosity of N$_2$ ice, the actual ice-shell thickness, and the possibility of a subsurface ocean. Nevertheless, they still offer important insight into the pressures and temperatures experienced at the impact site, and the displacement of the target mass. We validated our simulation by first considering a $2$ \kms\ projectile and the same target layering as \citet{Trowbridge2015}. We recover several qualitative aspects of their craters, including a central uplift when N$_2$ is added, steeper walls and deeper center compared to H$_2$O only, and the buildup of material at and beyond the crater rim.

\begin{figure}
  \includegraphics[width=0.5\textwidth]{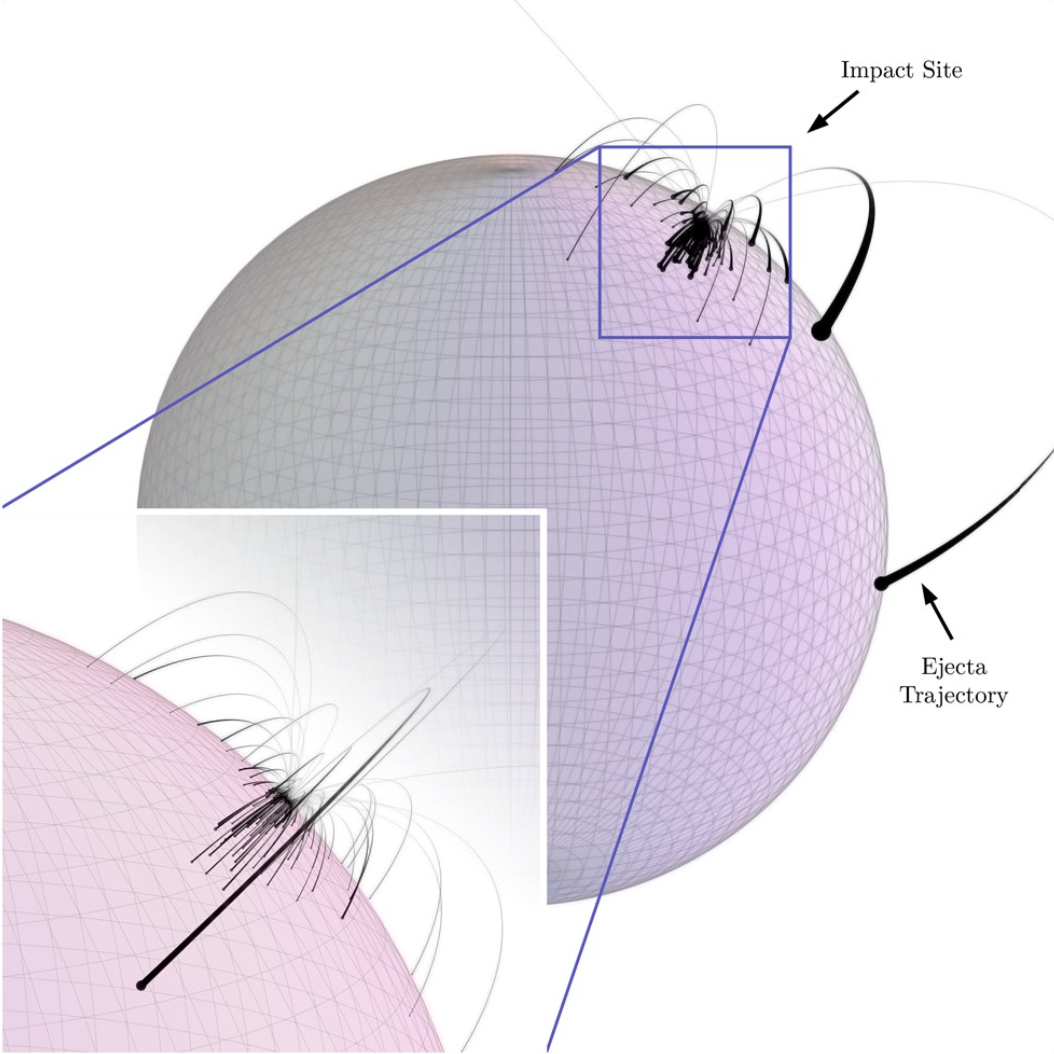}
  \caption{Trajectories of high-speed ejecta resulting from a KBO impact on Pluto. The paths of individual ejecta particles are determined from a \texttt{rebound} $N$-body simulation. The initial conditions for the particles were obtained from the output of the iSALE-2D simulations: a set of Lagrangian tracer particles which attained sufficient height with positive vertical velocity. For visualization the particles were given random horizontal orientation, which does not affect the distance traveled. The inset of the figure zooms in on the impact site to resolve more ejecta paths. Only $0.12\%$ of ejecta attained unbound trajectories and escaped Pluto.}
  \label{fig:ballistic}
\end{figure}

We proceeded to simulate a KBO impact on Pluto involving a $3.7$ \kms\ projectile, as discussed in the previous section, and maintained the $3\,\text{km}$ deep N$_2$ ice layer modeled by \citet{Trowbridge2015}. After 200 seconds, the modified crater exhibited a more complex floor morphology, which was depressed at the center. The steep walls were slightly terraced, and the buildup at the crater rim formed a scarp (left side of Figure~\ref{fig:impact}). The peak shock pressures, as recorded by a dense set of Lagrangian tracer particles extending 5 km deep and 10 km wide, agree within a factor of $\sim 2$ of those predicted by the  wave-interference model presented in \citet{Melosh1984}. The latter is shown as contours on the right side of Figure~\ref{fig:impact}. 

To further investigate the fate of high-speed ejecta, we filtered tracer particles to those which reached a height of at least 50 m while simultaneously having a positive vertical velocity component. Their phase-space coordinates were transferred to massless test particles in a \texttt{rebound} $N$-body simulation \citep{Rein2019}, centered on a massive body with the physical characteristics of Pluto. Horizontal velocities were re-oriented at random (without modifying their magnitudes) to simulate ballistic trajectories in three dimensions. It is important to note that this step has no effect on the distance that particles travel from the impact site. Figure~\ref{fig:ballistic} depicts some example trajectories for a subset of test particles. In total, about $0.12\%$ of the $\sim 8000$ particles from the excavated crater in iSALE-2D had sufficiently high velocity to escape, whereas the rest collided with Pluto. 

As an order-of-magnitude estimate of the corresponding mass, we note that approximately half of these tracer particles were classified as high-speed ejecta. The ratio of the mass excavated to the mass of the projectile was $\sim 100$, yielding approximately $0.06$ of the impactor's mass that escaped. This fraction is roughly $85\times$ larger than that predicted strictly from the wave-interference model, which might possibly be attributed to inaccuracies in the model beyond ejection speeds of 1 \kms\ \citep{Melosh1984}, the adopted material properties in the iSALE-2D simulation (e.g. the impactor and target surface comprise different materials), or the ejection of some material from beneath the spall layer \citep{Melosh1985}. Nevertheless, the coupled hydrodynamical and dynamical simulations confirm that only a minuscule portion of the excavated volume attains the velocity required to escape the KBO. This effect results in a downward revision of $f_{\text{ej}}$ in Equation \ref{eq:n2population} and a corresponding decrease in $N_{\text{iso}}$ in our Equation \ref{eq:drake}. 

Oblique impacts could be responsible for an enhancement of mass ejected from the KBO. Simulated $10$ \kms\ asteroid impacts on Mars indicate that up to $\sim 100\times$ more high-speed ejecta in oblique impacts than in vertical ones \citep{Artemieva2004}; however, this applies only to the narrow range in impact angle near $30^{\circ}$, with more typical enhancements of order $10-30\times$. At $35$ \kms, oblique comet impacts generate an enhancement only up to $\sim 12\times$. These factors do not provide sufficient revision to $f_{\text{ej}}$ to be compatible with the N$_2$ hypothesis.

\section{Observation of Exotic Ices} \label{sec:observation}

Since larger survey volumes should increase the known small-body population by an order-of-magnitude within the next decade \citep{ivezic2019lsst}, future \om-like objects may be identified before perihelion. Because these early detections should lead to comprehensive characterization with ground-based follow-up surveys, we discuss the potential observational signatures of outgassing from the candidate hydrogen and nitrogen ices.

\subsection{Production Rates and Coma Number Density}

\cite{Seligman2020} calculate `Oumuamua's mass evolution and sublimation rate demanded by the observed non-gravitational acceleration throughout the object's trajectory for a number of candidate ices. For this order-of-magnitude analysis, however, we follow Equation 4 from \cite{jackson20211i} to compute the magnitude of the non-gravitational force, $F_{\text{ng}}$, from outgassing of volatiles as

\begin{equation} \label{eq:nongrav}
    \lvert F_{\text{ng}} \rvert = \lvert M_{\text{iso}}a_{\text{ng}} \rvert \approx \frac{1}{3}\,\Bigg(\frac{dM_{\text{iso}}}{dt}\Bigg)\,v_{\text{jet}}\,,
\end{equation}

\noindent where $M_{\text{iso}}$ is the ISO mass, $a_{\text{ng}}$ is the magnitude of the non-gravitational acceleration, and the jet velocity is

\begin{equation} \label{eq:jetvelo}
    v_{\text{jet}} \approx \tau\, \sqrt{\frac{8kT_{\text{s}}}{\pi \mu m_{\text{H}}}}\,,
\end{equation}

\noindent where $T_{\text{s}}$ is the surface temperature of the ISO, $\mu$ is the molar mass of the outgassing particle, $m_{\text{H}} = 1\,\text{amu}$, and $\tau$ is related to an average of the outflow velocites and the Mach number of the jet. We use $\tau \approx 0.45$, as suggested by \citet{Crifo1987} and adopted in Section 2.3 of \citet{jackson20211i}. For an outgassing species $\text{x}$ with production rate $\text{Q(x)}$, we can write $(dM_{\text{iso}}/dt) = \text{Q(x)}\mu m_{\text{H}}$.

For the magnitude of the non-gravitational acceleration, we assume the model from \cite{micheli2018non}, where `Oumuamua's non-Keplerian trajectory depends on the inverse square of the heliocentric distance $r_{\text{h}}$:

\begin{equation} \label{eq:micheli}
    a_{\text{ng}} = A_{1}\,\Bigg(\frac{1\,\text{AU}}{r_{\text{h}}}\Bigg)^{2}\,.
\end{equation}

\noindent Here, $A_{1} = 4.9\,\times10^{-4}\,\text{cm}\,\text{s}^{-2}$ is the radial non-gravitational acceleration at $r_{\text{h}} = 1\,\text{AU}$, per the formalism of \cite{marsden1973comets}. From these equations, the production rate $\text{Q(x)}$ can be expressed as,

\begin{equation} \label{eq:production}
   \text{Q(x)} \approx \frac{3 A_{1}M_{\text{iso}}}{\mu m_{\text{H}}v_{\text{jet}}}\,
    \Bigg(\frac{1\,\text{AU}}{r_{\text{h}}}\Bigg)^{2}\,.
\end{equation}

In addition, the number density, $ \text{n(x)}$, of species $\text{x}$ in the coma of an actively-outgassing object can be calculated from Equation 1 in \cite{ootsubo2012akari} as

\begin{equation} \label{eq:coma}
    \text{n(x)} \approx \frac{\text{Q(x)}}{4\pi v_{\text{ex}}r_{\text{nuc}}^{2}}
    \exp{\,\Bigg(\frac{-r_{\text{nuc}}}{v_{\text{ex}}\tau_{x}}\Bigg)}\,,
\end{equation}

\noindent where $v_{\text{ex}}$ is the expansion velocity, $r_{\text{nuc}}$ is the nucleocentric distance, and $\tau_{\text{x}}$ is the photodissociation timescale of the species. For this analysis, we approximate that $v_{\text{ex}} \sim 10\, v_{\text{jet}}$. Observations of CO and other volatile species in the comet Hale-Bopp's outflow demonstrated that $v_{\text{ex}} \sim 1000\,\text{m/s}$ once the gas is warmed by solar radiation  (for example, see Figure 2 in \cite{biver1997long}), which will satisfy this relation within a factor of order unity for either \htwo{} or \ntwo{}. Substituting Equation \ref{eq:production} into Equation \ref{eq:coma}, the number density may be expressed as,

\begin{equation} \label{eq:finalcoma}
    \text{n(x)} \approx \frac{3A_{1}M_{\text{iso}}}{320\tau^2 kT_{\text{s}}r_{\text{nuc}}^{2}}
    \Bigg(\frac{1\,\text{AU}}{r_{\text{h}}}\Bigg)^{2}
    \exp\,\Bigg(\frac{-r_{\text{nuc}}}{10\,v_{\text{jet}}\tau_{x}}\Bigg)\,.
\end{equation}

Equation \ref{eq:coma} assumes a spherically-symmetric outflow, although it should still hold for lines-of-sight behind the nucleus. We will apply this simple coma model to the candidate exotic ices -- \htwo{} and \ntwo{} -- in order to predict the number density of outgassing volatiles that may be found in future observations.

\subsection{Hydrogen Outgassing}

For a hydrogen-propelled `Oumuamua at $1\,\text{AU}$, we find $v_{\text{jet}} \sim 113\,\text{m}\,\text{s}^{-1}$ and use $T_{\text{s}} = 6\,\text{K}$ \citep{Seligman2020}. With a photodissociation rate of $\tau_{\text{\htwo}} \sim 9\times10^{6}\,\text{s}$ for a quiet Sun at this $r_{\text{h}}$ \citep{huebner1992solar}, the e-folding length for molecules to be destroyed is of order $10^{7}\,\text{km}$. Thus, we can approximate the exponential term in Equation \ref{eq:finalcoma} to be unity for relatively small $r_{\text{nuc}}$. From these values, we can estimate the number density profile as

\begin{equation} \label{eq:comaH2}
    \text{n}\big(\text{\htwo}\big) = 1\times10^{5}\,\text{cm}^{-3}
    \Bigg(\frac{M_{\text{iso}}}{5\times10^{10}\,\text{g}}\Bigg)
    \Bigg(\frac{10^8\,\text{cm}}{r_{\text{nuc}}}\Bigg)^{2}\,,
\end{equation}

\noindent where we have based the fiducial mass on the estimate of \cite{Seligman2020} for the case where ISO density $\rho \sim 0.5\,\text{g}\,\text{cm}^{-3}$. We note that a pure hydrogen ice ISO would not reconcile the observed non-gravitational acceleration, as the surface coverage fraction required for an oblate ellipsoid scales as $0.06(\rho_{\text{iso}}/\rho_{\text{\htwo}})$ \citep{Seligman2020}. Objects denser than \htwo{} ice, as would be expected for heavily-weathered bodies, would require higher hydrogen surface coverage.

Although hydrogen ice outgassing would be undetectable at infrared and optical wavelengths, \htwo{} can dissociate via ultraviolet photons. For example, any of the following pathways may occur \citep{feldman2004spectroscopic}:

\begin{itemize}
    \item \htwo{} + $h\nu$ $\longrightarrow$ H + H $\;\;$ ($844.7\,\text{\AA}$)\,,
    \item \htwo{} + $h\nu$ $\longrightarrow$ \htwo$^{+}$ + e$^{-}$ $\;\;$ ($803.67\,\text{\AA}$)\,,
    \item \htwo{} + $h\nu$ $\longrightarrow$ H + H$^{+}$ + e$^{-}$ $\;\;$ ($695.8\,\text{\AA}$)\,.
\end{itemize}

\noindent For the above processes, we give the wavelength of the minimum possible energy photon in parentheses.

Emission from excited \htwo{} states has already been detected in Solar System small bodies by the \textit{Far Ultraviolet Spectroscopic Explorer} (FUSE), a NASA spacecraft which launched in 1999 and was operational for almost a decade, and had  wavelength coverage from  $905\text{\AA}-1187\text{\AA}$. Specifically, \cite{feldman2002far} attributed three lines ($1071.6\text{\AA}$, $1118.6\text{\AA}$, and $1166.8\text{\AA}$) from the comet C/2001 A2 (LINEAR) to fluorescence from solar Lyman-$\beta$ photons. While these lines were attributed to H$_2$O ice, a similar signature could appear from hydrogen iceberg ISOs.

While simply confirming the presence of \htwo{} ice in `Oumuamua-like objects would be a remarkable result, detailed studies of these ISOs could constrain their D/H ratios. Because these objects would necessarily form in dark, cold, optically-thick regions in GMCs, cloud products should incorporate matter that is otherwise difficult to detect. There may be non-trivial fractionation in the ISM \citep{kong2016deuterium}, but D/H ratios in actively-outgassing \htwo{} ice objects could illuminate the chemical conditions in the starless cores from which they were born.

In the absence of a mechanism to purify the \htwo{} ice, these hypothetical ISOs would collect and retain volatiles less tenuous than \htwo{} along with refractory substances. Nevertheless, neither micron-sized dust nor volatiles endemic to molecular clouds such as CO were found to emanate from `Oumuamua. This non-detection is still somewhat curious, even for the \htwo{} hypothesis, and it would imply that the ISOs outgassing was dominated by the exotic ice.

\subsection{Nitrogen Outgassing}

Similar to our calculation for hydrogen ice, we can determine the observational signature of a nitrogen coma at $1\,\text{AU}$. With $T_{\text{s}} = 25\,\text{K}$ \citep{Seligman2020}, we find jet velocity $61\,\text{m}\,\text{s}^{-1}$. This $v_{\text{jet}}$ is slightly lower than what was published by \cite{jackson20211i}, although the \texttt{python} code provided with the manuscript returns this $61\,\text{m}\,\text{s}^{-1}$. We use $\tau_{\text{\ntwo}} \sim 7\times10^{5}\,\text{s}$ for a quiet Sun \citep{huebner1992solar}, which gives an e-folding distance for photodissociation of $v_{\text{ex}}\tau_{\text{\ntwo}} \sim 4\times10^{5}\,\text{km}$. Thus, we again set the exponential term in Equation \ref{eq:finalcoma} to unity and estimate the number density profile as 

\begin{equation} \label{eq:comaN2}
    \text{n}\big(\text{\ntwo}\big) = 7\times10^{3}\,\text{cm}^{-3}
    \Bigg(\frac{M_{\text{iso}}}{10^{10}\,\text{g}}\Bigg)
    \Bigg(\frac{10^8\,\text{cm}}{r_{\text{nuc}}}\Bigg)^{2}\,,
\end{equation}

\noindent for small nucleocentric distance $r_{\text{nuc}} \ll v_{\text{jet}}\tau_{\text{\ntwo}}$.

Implicit in this fiducial size is the predicted albedo of $0.65$ from \cite{jackson20211i}, and the study considers Pluto's overall albedo to represent a good match for `Oumuamua. However, `Oumuamua-sized nitrogen fragments would lose around 90\% of their mass during the Solar encounter \citep{jackson20211i}; this sublimation should uncover fresh ice with a potentially higher albedo. Instead of considering the weathered terrain covering most of Pluto and Triton as the best comparison, the relatively-young, nearly pure \ntwo{} terrain of Sputnik Planitia might be more in-line with expectations for a nitrogen iceberg. Unlike the predicted albedo of $0.65$ from \cite{desch20211i} and the reddish observed color of \om{} \citep{Meech2017}, however, Sputnik Planitia has an albedo of nearly unity and a bluish color \citep{olkin2017global}. Thus, a nitrogen fragment should match the color of Pluto in interstellar space but not necessarily after its strongest outgassing activity surrounding perihelion.

As with \htwo, we expect \ntwo{} outgassing to be detectable at ultraviolet wavelengths \citep{jackson20211i}. Common photodissociation reactions for \ntwo{} include \citep{huebner1992solar} (and references therein):

\begin{itemize}
    \item \ntwo{} + $h\nu$ $\longrightarrow$ N + N $\;\;$ ($1270.4\,\text{\AA}$)\,,
    \item \ntwo{} + $h\nu$ $\longrightarrow$ \ntwo$^{+}$ + e$^{-}$ $\;\;$ ($796\,\text{\AA}$)\,,
    \item \ntwo{} + $h\nu$ $\longrightarrow$ N+ N$^{+}$ + e$^{-}$ $\;\;$ ($510.4\,\text{\AA}$)\,.
\end{itemize}

In Solar System comets, attempts by the \textit{FUSE} spacecraft to identify molecular nitrogen outgassing resulted in non-detections \citep{bockelee2004composition}. Because \ntwo{} is one of the least spectroscopically-active nitrogen-bearing species \citep{cochran2000n+}, searches for this constituent have focused on the weak N$_{2}^{+}$ (0, 0) band at $3914\,\text{\AA}$. Furthermore, space-based observatories seem better suited to search for \ntwo{} in comets \citep{feldman2004spectroscopic}, given the plethora of molecular nitrogen in the Earth's atmosphere. Nonetheless, this feature was unambiguously identified from ground-based telescopes in the comet C/2016 R2 \citep{opitom2019n2, mousis2021cold}. Moreover, the object's unusual composition led \cite{desch20211i} to suggest that it may be a collisional fragment from our own solar system. While not an interstellar object, the formation pathway would be consistent with that the \ntwo{} hypothesis for `Oumuamua.

Unlike `Oumuamua, however, a bright visible coma and strong CO emission were also detected from C/2016 R2. Thus, these objects could not be directly analogous, since `Oumuamua would require nearly pure \ntwo{} to match the observational constraints. If `Oumuamua-like ISOs have \ntwo{} as a dominant constituent, the aforementioned $3914\,\text{\AA}$ feature could be discoverable in future interlopers.

\section{Non-Exotic Ice Compositions} \label{sec:nonice}

Tenuous, exotic ices can naturally explain \om's elongated geometry as an oblate ellipsoid resulting from weathering in the ISM. However, other interpretations may also reconcile the non-Keplerian trajectory with a lack of common volatiles via generic objects of either the disk-like or cigar-like shape.

\subsection{Ultra-Porous Dust Aggregates}

Solar radiation pressure was originally discussed and dismissed as important for `Oumuamua \citep{micheli2018non} because the magnitude of non-gravitational acceleration would require a bulk density orders-of-magnitude lower than any other known small body. Nonetheless, \cite{bialy2018could} revisited this hypothesis and found that `Oumuamua could be propelled by this phenomenon if its bulk density were $\rho \sim 10^{-5}\,\text{g}\,\text{cm}^{-3}$ or if it were very thin. \cite{moro2019could} examined the feasibility of building the ultra-porous diffusion-limited aggregates from grains in protoplanetary disks, and \cite{luu2020oumuamua} proposed that a similar physical process may occur in the tails of fragmented comets. The former formation mechanism has no preference for `Oumuamua's geometry, while the latter would result in less-favored, but still-allowed prolate geometry.

The mechanical stability of these dust aggregates subjected to tidal torques around perihelion was demonstrated by \cite{flekkoy2019interstellar}. Furthermore, \cite{seligman2021CO} demonstrated that the spin-state and thus, the lightcurve, of objects driven by solar radiation pressure is stable. Moreover, the creation of extremely-porous aggregates would not deplete the mass reservoir of rocky and icy material in extrasolar systems. With $N_{\text{iso}} \sim 10^{16}$ and applying Equation \ref{eq:diskscaling} from Section \ref{sec:oom}, populating the galaxy with these objects requires roughly $7\times10^{-5}\,\text{\mearth}$ for the fiducial size $r_{\text{iso}} \sim 100\,\text{m}$. This result introduces no tension with the mass budget of protoplanetary disks.

While this compatibility holds if objects are directly-assembled from micron-sized dust grains, the proposed mechanism of \cite{luu2020oumuamua} requires a decameter-sized comet fragment to host the growing aggregate. Therefore, the theoretical upper bound on number density for these objects would be driven by the mass reservoir of the intermediary $10\,\text{m}$ boulders. With boulders of density $0.6\,\text{g}\,\text{cm}^{-3}$, then each protoplanetary disk must devote $4\times10^{-3}\,\text{\mearth}$ of material to this formation pathway as determined from Equation \ref{eq:diskscaling}. We have fiducially assumed that each boulder forms one ISO. Therefore, even this more mass-intensive channel could still easily satisfy the inferred ISO reservoir.

Despite the reconciliation with the required mass reservoir, there is no proposed or apparent destructive mechanism for these dust aggregates in the ISM. This implies that there should be no preference for kinematically-young objects like `Oumuamua. If the formation of these aggregates is widespread throughout the galaxy, then `Oumuamua's LSR velocity would be anomalously low. Interestingly, the dust aggregate hypothesis does allow for a single protostellar disk in COL/CAR to populate the local region with $n_{\text{iso}} \sim 0.1\,\text{AU}^{-3}$ by ejecting of order $1\text{\mearth}$ of porous bodies. Thus, `Oumuamua's kinematic age would suggest an origin from an abnormal protoplanetary disk in COL/CAR as opposed to an undiscovered, yet common astrophysical process. This interpretation would corroborate the conclusion from \cite{moro2018originI, moro2018originII} that `Oumuamua could not be a member of an isotropic background population if it originated from an extrasolar system. This scenario's mass budget requires the direct assembly from \cite{moro2019could} as opposed to the fragment-mediated pathway from \cite{luu2020oumuamua}.

If the grains in ultra-porous dust aggregates consist mostly of cometary volatiles as presented by \cite{moro2019could} and \cite{luu2020oumuamua}, then the ISOs would be subject to destruction during the Solar encounter. Taking water ice as an example, we find that all of the object's mass could be lost within 1 hour at $1\,\text{AU}$. As `Oumuamua received more than $10^{12}\,\text{erg}\,\text{cm}^{-2}$ of Solar flux while it was inside $2\,\text{AU}$ \citep{Seligman2020}, the ISO would require substantial refractory substances to retain its structural integrity. This complete dessication, however, could increase the porosity and decrease the density, making propulsion by solar radiation pressure more effective.

\subsection{A Solar System Small Body Analogue}

\cite{seligman2021CO} showed that carbon monoxide (CO) outgassing could reconcile the energetics of the non-gravitational acceleration with outgassing. While this solar system volatile could place the ISO within known realms of small bodies, the \textit{Spitzer} non-detection places strict limits on cometary activity from this substance during the observational window. Because the images were collected only during a 33 hour time frame, \citep{seligman2021CO} suggested that `Oumuamua could have experienced sporadic jets and been in a low-activity state during the attempted \textit{Spitzer} observation.

An explanation of `Oumuamua as a solar system analogue would be a priori compelling, as these comparatively benign objects were expected as interstellar interlopers before Pan-STARRS. If this ISO were a comet-like body, however, its inbound kinematics versus the LSR would imply an unexpectedly-low age. Furthermore, `Oumuamua's extreme aspect ratio would still make it an anomaly even if its bulk composition is not exotic. Population-level statistics will help constrain small body formation processes in extrasolar systems and determine whether this interpretation is viable.

\subsection{A Solar System Origin}

Initially, it was speculated that `Oumuamua may have been a Solar System object that was scattered onto a hyperbolic trajectory. However, \citet{Schneider2017} found that this was unlikely due to the ISO's inbound direction from the galactic apex. Moreover, its orbital energy was too large to be the result from a single scattering event with any of the known or hypothetical giant planets \citep{Wright2017}. While \citet{delafuente2018} hypothesized that `Oumuamua may have been a perturbed Oort cloud object, all of these interpretations involving an origin in our own solar system would necessitate `Oumuamua's physical attributes to be an extreme outlier among the population of known solar system small bodies of similar size.

\subsection{An Artificial Object}

\cite{bialy2018could} suggested that \om{} could be artificial, since no natural explanation has been perfectly satisfactory. Due to the uncertainty in the prevalence of such bodies, we cannot propose reasonable estimates for any parameters in Equation \ref{eq:drake}. Nevertheless, the theoretical stability of any ellipsoidal object's spin state under radiation pressure, including a lightsail, was demonstrated by \cite{seligman2021CO}. 

One could posit that if `Oumuamua were artificial, then attempts to detect radio signals from the object would have been successful. However, observations with the Robert C. Byrd Green Bank Telescope \citep{Enriquez2018}, the Murchison Widefield Array \citep{Tingay2018} and the Allen Telescope Array \citep{Harp2019} resulted in non-detections. %Moreover, there has yet to be any evidence pointing specifically towards a non-natural explanation for `Oumuamua \cite{oumuamua2019natural}.

\subsection{Other Proposed Interpretations for `Oumuamua}

To explain the elongated geometry, \citet{Hansen2017}, \citet{Rafikov2018b}, and \cite{katz2018interstellar} offered hypotheses that `Oumuamua's birthplace was a post-main sequence star system. As an alternative, \citet{Jackson2017} and \citet{Cuk2017} both proposed that `Oumuamua was a rocky body ejected from a circumbinary system. However, the later discovery of `Oumuamua's non-ballistic trajectory has largely ruled-out these interpretations.

While non-gravitational effects have been detected in active asteroids \citep{hui2017active}, the magnitudes of the resulting accelerations are significantly smaller for these objects than for `Oumuamua. Although the magnitude of `Oumuamua's non-gravitational acceleration is among the highest when compared to those of solar system small bodies, its physical size is also among the smallest of the population for which this effect has been measured. Thus, this relatively high value for `Oumuamua should be expected if its composition were similar to these previously observed objects. Members of the Manx \citep{meech2016manx} and Damacloid \citep{jewitt2005damacloids} groups, although they are unusual objects for their locations in the solar system, could not be analogues for `Oumuamua. Their low activity levels preclude them from attaining the non-Keplerian trajectory demanded for generic `Oumuamua-like objects with typical cometary volatiles, and no exotic ices have been found.

\citet{raymond2018implications} proposed that \om{} could be a fragment from a tidally-destroyed small body, and \cite{zhang2020tidal} demonstrated that this process would generally create a prolate geometry. Because typical cometary volatiles would fuel the non-gravitational acceleration, this formation mechanism requires the progenitors to be abundant in CO or other energetically-efficient propellants. However, the population constraints as described in Section \ref{sec:oom} for any extrasolar system product would demand that a substantial fraction of the ejecta be converted to these tidally-disrupted ISOs.

\section{Discussion} \label{sec:discussion}

\subsection{Occurrence of Borisov-Like Objects}

While the nature of `Oumuamua remains unresolved, C/2019 Q4 (later renamed 2I/Borisov) was identified\footnote{Minor Planet Electronic Circular (MPEC) \href{https://www.minorplanetcenter.net/mpec/K19/K19RA6.html}{ 2019-R106}} in 2019 and immediately determined to be a cometary body. This second ISO -- with inbound velocity $v_{\infty} = 32.3\,\text{km}\,\text{s}^{-1}$ and eccentricity $e = 3.35$\footnote{JPL Horizons: \href{https://ssd.jpl.nasa.gov/sbdb.cgi?sstr=2I;old=0;orb=0;cov=0;log=0;cad=0\#elem}{C/2019 Q4} (accessed 05 Apr 2021)} -- was discovered before perihelion, allowing for detailed studies of its composition and evolution during the Solar encounter.

Unlike `Oumuamua, Borisov displayed a brilliant visible coma. Optical images with the \textit{Hubble Space Telescope} provided evidence for a nucleus size of $200-500\,\text{m}$ \citep{Jewitt2019b}. While H$_2$O sublimation drove activity starting at $6\,\text{AU}$ \citep{Fitzsimmons:2019, Jewitt2019b, Ye:2019}, carbon-bearing and nitrogen-bearing species were also detected in the outflow throughout the ISO's trajectory \citep{Opitom:2019-borisov, Kareta:2019, Lin:2019,Bannister2020}. Notably, the coma was rich in CO compared to typical comets in the solar system \citep{Bodewits2020, Cordiner2020}. Dust in the cometary tail was found to have a color of $g'-r' = 0.63 \pm 0.03$ \citep{Guzik:2020,Jewitt2019b, Bolin2019}, and there was a paucity of micron-size dust. Instead, most grains were larger than $100\,\mu \text{m}$ \citep{Kim2020}. %\citep{Bagnulo2021} measured polarimetry in dust

In contrast with `Oumuamua, Borisov was far brighter than the apparent magnitude limit for surveys such as Pan-STARRS. Thus, its observation does not as tightly constrain the galactic population of its class. Because pathways of traditional small body formation are better-known \citep{Chiang2010,Morbidelli2016} compared to those of \om-like ISOs, the population of Borisov-like objects may be best-derived from theoretical estimates until VRO determines tighter observational constraints.

Gravitational focusing can increase the occurrence of stellar encounters for young ISOs such as `Oumuamua over those such as Borisov. We quantify the impact parameter $b_{\text{iso}}$ per \cite{raymond2018implications} as 

\begin{equation}
    b_{\mathrm{iso}}^{2} = q^{2} \Big(1 + \frac{v_{\mathrm{esc}}^{2}}{v_\mathrm{\infty}^{2}} \Big)\,,
\end{equation}

\noindent where $v_{\infty}$ is the inbound ISO velocity relative to the Solar System and $v_{\text{esc}}$ is the  escape velocity from the Sun at perihelion distance $q$. 

Although the nature and velocity distribution of `Oumuamua-like objects is unknown, we take this first-detected member as representative. Thus, we use $v_{\infty} = 26\,\text{km}\,\text{s}^{-1}$ and $q = 0.25\,\text{AU}$ to find $b \approx 11q$ for these young ISOs. As a comparison, Borisov had impact parameter $b \approx 2q$. Due to the Sun's velocity versus the LSR drawn from the dispersion of Population I stars, it is improbable for a given ISO to match the Sun's galactic velocity. Nonetheless, gravitational focusing can enhance the observation rate of kinematically-young objects by a factor of a few. In the future, the typical impact parameter for ISOs will be determined by telescopes such as VRO.

Because Borisov-like small bodies originate in extrasolar systems, some proposed formation scenarios for `Oumuamua-like objects would then require that the protoplanetary disk's mass reservoir be split between building these two types of ISO. In our previous analysis on the theoretical number density of disk product hypotheses, we have assumed that `Oumuamua-like objects can harbor the entire available mass. Given Borisov's size and impact parameter, the mass density of these bodies is far lower than what's inferred for `Oumuamua-like objects. Therefore, it seems as though more of the ejecta should be small objects without visible coma. While the lack of identifications of other Borisov-like objects is interesting, it does not create tension with constraints on `Oumuamua-like objects. In reality, both `Oumuamua and Borisov are members of a size distribution of ISOs \citep{Bolin2020AJ}, and future surveys may discover a diverse population.

\subsection{Expectations for the Vera Rubin Observatory}

The forthcoming VRO/LSST will provide narrower constraints on the prevalence of interstellar objects. Provided that the discoveries of \om{} and Borisov were indicative of galactic reservoirs, then VRO should provide population-level statistics by detecting objects up to three magnitudes dimmer than Pan-STARRS \citep{trilling2017implications}. Furthermore, it will identify ISOs early enough in their trajectories through the Solar System to permit extensive follow-up observations. Within the next decade, the timely detection of an ISO may lead to an in-situ mission such as the ESA's \textit{Comet Interceptor} \citep{snodgrass2019european}, or the NASA concept study \textit{BRIDGE} \citep{Moore2021}. Visiting one `Oumuamua-like object would provide unparalleled detail on material that originated outside of our Solar System \citep{Seligman2018}.

Table \ref{tab:VRO} outlines the predicted number of `Oumuamua-like objects  to be identified by VRO, $N_{\text{VRO}}$, as a consequence of the various interpretations. In addition, we discuss the number objects $N_{\text{young}}$ with age less than $1\,\text{Gyr}$. For each interpretation, we assume that the ISOs have number density equal to the minimum of our fiducial $n_{\text{iso}}$ as inferred from Pan-STARRS and the theoretical upper limit from their formation pathway. Simply from the occurrence rate of similar ISOs in the future, VRO could differentiate between the proposed interpretations for the first interstellar visitor.

If `Oumuamua was a protoplanetary disk product, then future surveys will be critical to determining the prevalence of these objects which we would consider anomalous in our Solar System. By placing `Oumuamua in the context of other ISOs, we can place our Solar System in the context of exoplanetary systems. The mass reservoir of protoplanetary disks can satisfy the inferred $n_{\text{iso}}$, so $N_{\text{VRO}}$ should be similar to the prediction from the VRO/LSST team \citep{ivezic2019lsst}.

While hydrogen icebergs face perhaps insurmountable temperature barriers to formation, starless cores could provide a sufficiently large mass reservoir to supply the local Solar neighborhood with the inferred population of ISOs. Because of the transient nature of these objects, any future \htwo{} icebergs should mirror `Oumuamua's low LSR velocity. If this hypothesis is correct, it would suggest that any other objects with \htwo{} activity would most likely arrive from COL/CAR in order to reconcile constraints on the galactic population of their remnants. In contrast, the theoretical bounds on the reservoir of \ntwo{} icebergs dictate that observing them should be rare. Notwithstanding the issues surrounding the cratering process' ability to excavate \ntwo{} ice from exo-Plutos, the widespread formation of these collisional shards would still not predict the detection of one by Pan-STARRS. Even with a decade-long survey by VRO, it would be unlikely to detect another nitrogen iceberg. However, \cite{desch20211i} suggest that the population of H$_2$O shards could be nearly an order-of-magnitude higher than those composed of N$_2$. If these collisional byproducts exist, then VRO may detect a water ice fragment.

VRO/LSST is also expected to expand the population of identified Solar System small bodies by at least tenfold \citep{ivezic2019lsst}. Because \cite{desch20211i} predicted that 0.1\% of long-period comets gravitationally-bound to the Sun could be collisional fragments from our Kuiper Belt, a na\"ive extrapolation suggests that VRO/LSST should identify at least several of these objects. While the aforementioned nitrogen-rich C/2016 R2 was mentioned as a candidate for this class of object by \cite{desch20211i}, its spectrum was starkly different from that of `Oumuamua \citep{mousis2021cold}. If `Oumuamua-like objects are \ntwo{} icebergs, then the expanding small body catalog may begin to include more direct analogues originating in our solar system.

\begin{table}[]
    \centering
    \begin{tabular}{c|c|c}
        Interpretation & $N_{\text{VRO}}$ & $N_{\text{young}}$ \\
        \hline
        Solar System Analogues & 2-30 & 0-3\\
        Hydrogen Icebergs & 2-30 & 2-30\\
        Nitrogen Icebergs & 0-1 & 0-1\\
        Galactic Dust Aggregates & 2-30 & 0-3\\
        COL/CAR Dust Aggregates & 2-30 & 2-30\\
    \end{tabular}
    \caption{Number of `Oumuamua-like objects predicted to be found by VRO/LSST, $N_{\text{VRO}}$, based on the results of our study. In addition, we predict the number of ISOs to be detected $N_{\text{young}}$ with an object age of less than $1\,\text{Gyr}$. We assume an expectation value of $1\,\text{yr}^{-1}$ for a 10 year survey from \cite{ivezic2019lsst}, then apply the results of our number density analysis with the $1\sigma$ confidence interval.}
    \label{tab:VRO}
\end{table}

If `Oumuamua was a dust aggregate, then the small mass budget required for a widespread galactic process would suggest that VRO will find similar objects -- albeit with a wide inbound velocity distribution. Because Pan-STARRS would have seen one such object, there would be no \textit{a priori} reason that observing an ultra-porous object would be a fluke. Thus, $N_{\text{VRO}}$ should be similar to previous predictions based on the identification of `Oumuamua. In the scenario where these ISOs originated in one atypical protoplanetary disk from COL/CAR, then $N_{\text{VRO}} \sim N_{\text{young}}$.

\om-like ISOs may represent samples of frigid interstellar environments, extrasolar systems with divergent evolutionary tracks, planetary systems at different stages in their life cycle, or analogues of regions of our Solar System that are difficult to access. Although the window to gather data from `Oumuamua has passed, deeper surveys such as VRO/LSST will permit  detailed follow-up of future interlopers. Regardless of the final verdict on their composition, these interstellar objects will provide insight about previously-unconsidered astrophysical processes.

\vspace{5mm}

\acknowledgments
We thank the two anonymous reviewers for their insightful reports that improved the manuscript. Additionally, we thank Karen Meech, Marco Micheli, Davide Farnocchia, Robert Jedicke, Karin \"Oberg, Mark Walker, Christopher Lindsay, Emma Louden, and Adina Feinstein for the useful discussions. Our section on hypervelocity impacts was improved thanks to insightful comments from Brandon Johnson. We gratefully acknowledge the developers of iSALE-2D, including Gareth Collins, Kai Wünnemann, Dirk Elbeshausen, Tom Davison, Boris Ivanov and Jay Melosh.

\pagebreak
\bibliography{bibliography}
\bibliographystyle{aasjournal}

\end{document}